\documentclass[twocolumn,aps,prl,showpacs]{revtex4}
\usepackage{graphicx}
\begin{document}
\def\la{\langle}\def\ra{\rangle}
\def\vep{\varepsilon}
\def\wh{\widehat}
\newcommand{\beq}{\begin{equation}}
\newcommand{\eeq}{\end{equation}}
\newcommand{\beqa}{\begin{eqnarray}}
\newcommand{\eeqa}{\end{eqnarray}}
\title{Quantum time-of-flight measurements:
kick clock versus continuous clock.}
\author{Daniel Alonso}
\affiliation{Departamento de F\'\i sica Fundamental y Experimental,
\\Electr\'onica y Sistemas.Universidad de La Laguna,
La Laguna, Tenerife, Spain}
\author{R. Sala Mayato}
\affiliation{Departamento de F\'\i sica Fundamental II.
Universidad de La Laguna, La Laguna, Tenerife, Spain}
\author{J. G. Muga}
\affiliation{Departamento de Qu\'\i mica-F\'\i sica,
Universidad del Pa\'\i s Vasco, Apdo. 644, 48080 Bilbao, Spain}

\begin{abstract}
The measurement of time durations or instants of ocurrence of 
events has been frequently modelled  ``operationally'' by  
coupling the system of interest to a ``clock''.  
According to several of these models the operational approach is 
limited at low energies because the perturbation of the clock 
does not allow to reproduce accurately 
the corresponding 
ideal time quantity, defined for the system in isolation.
We show that, for a time-of-flight 
measurement model that can be set to measure dwell or arrival times,
these limitations may be overcome by  
extending the range of energies where the clock works 
properly  using pulsed couplings 
rather than continuous ones. 
\end{abstract}
\maketitle


In the standard formulation of quantum mechanics the measurements 
are assumed to take place at one instant of time. Correspondingly, the
expectation values are calculated for particular instants. Within this 
theoretical scheme it is
not obvious how to obtain averages or, more generally, 
distributions for durations of processes,
or for the instants when a particular event
occurs \cite{ML00,MSE02}. 
The approaches to formalize time observables
may be classified in two groups: those which rely on the system 
of interest alone to produce ``ideal quantities''; and ``operational
approaches'' where the system of interest is coupled to an auxiliary system 
or additional degree(s) of freedom which act as a clock or stopwatch.
The coupling and auxiliary system may be more or less realistic, but
even simple ``toy models'' have been useful to gain insight
into the peculiarities of time measurements. Several of these 
models show the limitations of the 
operational route to reproduce accurately either the corresponding 
ideal time quantity, or the reference of the classical time
for low particle energies, 
due to the perturbation induced by the clock
\cite{Peres80,Peres95,AOPRU98}.     
In this letter we show 
that these limitations may be overcome by  
extending the range of energies where the clock works 
properly  using pulsed couplings 
rather than continuous ones. 
Our study case is the time-of-flight gedanken experiment put forward 
by Peres \cite{Peres80}, based on the Salecker-Wigner
clock \cite{SalWig58}. With a minor modification this is essentially 
the same model used in \cite{AOPRU98} for a time-of-arrival measurement.

Consider first the clock in isolation, described by  
the Hamiltonian \cite{SalWig58}
$\wh H_c=\omega \wh J=-i \hbar \omega \partial_{\theta}$,
where $\wh J=-i\hbar\partial/\partial \theta$, 
$\omega$ is the angular frequency and $\theta \in [0, 2\pi)$
the angle.   
This Hamiltonian has normalized eigenfunctions
\beq
u_n(\theta)=\exp(i n \theta)/(2\pi)^{1/2},
\eeq
where $n$ is any integer.  
Let us choose now the following initial state,  
\begin{equation}
v(\theta, t=0)= \sum_n u_n(\theta)/\sqrt{N},
\end{equation}
where $-j\le n\le j$ and $N\equiv 2j+1$. 
$v(\theta,t=0)$ is peaked at $\theta=0$ with 
an ``angle uncertainty'' $2\pi/N$ (This is defined as the angle  
necessary to obtain an orthogonal state by rigid displacement, see
also Eq. (\ref{ta}) and the discussion below.)   
Since the evolution of this function with $\exp(-i\wh{H}_c t/\hbar)$
during a time $t$ is
simply a rigid diplacement along $\theta$ by $\omega t$, 
\beq
v(\theta, t)=v_0(\theta-\omega t,0),   
\eeq
the state acts as the hand of the clock. 
Between 0 and $2\pi$ its peak indicates the time 
(between $t=0$ and $t=2\pi/\omega$) as
$t=\theta_{peak}/\omega$ . Alternatively,
\beq
\label{tcont}
t=\la\wh{\theta}\ra/\omega.
\eeq
A single measurement of $\theta$ does not provide
$\la \wh\theta\ra$ or $\theta_{peak}$
because of the hand width, so it is useful to 
define
\beq\label{ta}
\tau=\frac{2 \pi}{N\omega}
\eeq
as the time resolution of the clock for a given hand
and frequency $\omega$. This is the time required to clearly 
separate two successive hand positions; more precisely,  
it is the time required to make two states orthogonal.   
Note that, at variance with ref. \cite{Peres80},
we do not restrict the Hilbert space of
the possible clock states by a finite basis set, so that 
$\theta$ or functions of $\theta$ may be considered 
observables here. A consequence is that there is no need 
to introduce a discretized clock time operator whose average provides 
the correct time only at multiples of $\tau$, nor there is any 
need to perform a calibration to improve its performance
\cite{Leavens93,LMK94}.
Eq. (\ref{tcont}) provides the correct parametric
time $t$ at all times, modulo $2\pi/\omega$.   

Let us next consider a particle  
of mass $m$ moving freely in one dimension 
with Hamiltonian 
$\wh H_s={\wh p_x^2}/{2m}$. In a certain region of length $d$
the particle is coupled to the  
clock as follows \cite{Peres80,Leavens93,LMK94,LS98,rae},    
\begin{equation}\label{contclock}\wh H=\wh p_x^2/2m +\chi_d(\wh{x})
\omega \wh J=\wh p_x^2/2m -i \omega\hbar \chi_d(\wh{x})
\partial /\partial \theta,
\end{equation}
where
\begin{equation}\chi_d(x)=
\cases{ 1, \,\, \hbox{if x}\in[0,d] \cr 0,\,\,\hbox{elsewhere}}
\end{equation}
is the characteristic function indicating that the clock
runs only when the particle ``is'' within the interval $[0,d]$. 

Noting that $\wh{H}$ commutes with $\wh J$, the time dependent
wave function with initial state 
$\psi(x)v(\theta,t=0)$
is given by  
\beq
\frac{1}{N^{1/2}}\sum_n\psi_n(x,t)e^{-in\omega t/\hbar}u_n(\theta),
\eeq
where $\psi_n(x,t)$ is the (partial) wave that evolves from 
$\psi(x)$ with the Hamiltonian 
\beq
\wh{H}_s+\wh{V}_n=\wh{p}_x^2/2m+n\hbar\omega \chi_d(\wh{x}),
\eeq
which represents a particle 
that collides with a rectangular barrier (well) of heigth 
(depth) $V_n=n \hbar \omega$ and width $d$.
For an incident particle
with energy $E=p_x^2/ 2m$ and wavenumber $k=\sqrt{2 m E}/\hbar$, the wavenumber 
inside the barrier is  $k' =\sqrt{2 m(E-V_n)}/\hbar$
so that the phase shift caused by the barrier is approximately
given by
\begin{equation}\label{ps}(k'-k)d \sim -n \omega t_f,\end{equation}
where $t_f=d/(2E/m)^{1/2}$ is the classical time of flight. 
The right hand side in  Eq. (\ref{ps}) is a good
approximation if $E >>|V_n|$,
which means that the disturbance caused by the measurement is negligible.
If the incident wave packet is very much peaked (in wave number) 
around
\begin{equation}\psi_i(x,\theta)=
e^{i kx} v_0(\theta)= e^{ikx} \sum_{n=-j}^j u_n(\theta)/\sqrt{2j+1},
\end{equation}
the outgoing one after the barrier for the particle plus the
clock system will be very much peaked around
\begin{equation}
\psi_f(x,\theta)\sim e^{i k x} v(\theta-\omega t_f(k)),
\end{equation}
so that the hand points at 
the expected classical time-of-flight through the region of
interest. In the clock, $|V_n|$ can be as large as
$j \hbar \omega \sim \pi \hbar / \tau$ so that the condition
of negligible disturbance is given by
\cite{Peres80,Peres95}
\begin{equation}\label{peres}\tau >>
\frac{\pi \hbar}{E} \,\,\,\, \hbox{or}\,\,\,\,
E >> \frac{\pi \hbar}{\tau}.
\end{equation}
The same result may be obtained by imposing that the transmission 
probability should be close to one \cite{AOPRU98}.

Eq. (\ref{peres}) imposes a lower limit on the time resolution
of the clock. Equivalently, it 
imposes a lower bound on the incident energy
of the particle such that a clock with resolution
$\tau$ can be considered as a small disturbance to the incident particle
during the measuring process.
It is worth noticing that this limitation affects equally 
the measurement of dwell times in the region of lenght $d$ 
\cite{Peres80},
or arrival times at $x=d$ \cite{AOPRU98},   
which are obtained respectively by locating the initial 
wave packet outside or 
inside the selected interval $[0,d]$.

Let us now work out a pulsed version of the particle-clock 
system to avoid the excessive disturbance of the continuous clock.    
The simplest realization of a pulsed interaction is a succession of 
instantaneous kicks separated by a time ${\sf T}$. 
(For examples of kicked systems see   \cite{CC95,SZAC2000a,SZAC2000b,SZAC2001}.
An experimental realization of  a {\it kicked rotor}
can be found in \cite{raizen}.
We refer the interested reader to \cite{CC95} for further 
details.) The Hamiltonian for the kicked Peres-Salecker-Wigner 
clock is \cite{rae} 
\begin{equation}\label{Hsw}\wh H(t) =\frac{\wh p_x^2}{2m}
+ {\sf T}  \delta_{\sf T}(t) \omega \chi_d(\wh{x}) \wh J,
\end{equation}
where we have defined
$\delta_{\sf T}(t)=\sum_{n=-\infty}^{\infty} \delta(n{\sf T}-t)$, 
and the evolution operator of the kicked clock right before two 
kicks is
\begin{equation}
\label{ukicksw}
\wh U^{\sf T}= e^{-\frac{i}{\hbar}{\sf T}
\wh p_x^2/2m} e^{- {\sf T} \omega \chi_d(x)
\partial /\partial \theta}=\wh U^{\sf T}_s \wh U^{\sf T}_{c-s}.
\end{equation}
If ${\sf T}\to 0$,
the infinitesimal evolution operator,
$\wh U^{{\sf T} \to 0}$, of the kicked system coincides
with the infinitesimal evolution operator of the continuous-coupling clock 
up to ${\cal{O}}({\sf T}^2
[\wh H_s,\wh H_{c-s}])$, which goes to zero with ${\sf T}$.
The difference between the pulsed and continuous-coupling clocks will
therefore be seen for 
larger values of ${\sf T}$. These larger values may also avoid an excessive 
perturbation, but the time interval
between kicks cannot be arbitrarily large. If we want 
to extract a characteristic time scale of the particle, ${\sf T}$ must be  
smaller than the time scale we want to measure.

We shall next show that  
the kicked clock can be succesfully used for energies 
which are smaller than $\pi \hbar/\tau$. Note first that 
\begin{equation}\label{U1}\wh U^{\sf T}_{c-s} u_n(\theta)=
\exp(-i {\sf T} \omega n) u_n(\theta).
\end{equation}
If the phase, ${\sf T} \omega n$, is written as
\begin{equation}{\sf T} \omega n =2 \pi s + 
\nu_n , \,\, s \in {\cal N},\,\, \nu_n \in {\cal R}\end{equation}
then
\begin{equation}
\label{as2}\hbar \omega n=
\frac{2 \pi \hbar}{\sf T}s+\frac{\hbar}{\sf T} \nu_n,
\end{equation}
and Eq. (\ref{U1}) takes the form
\begin{equation}
\wh U^{\sf T}_{c-s} u_n(\theta)  = \exp(-i \nu_n) u_n(\theta).
\end{equation}
For the different energies of the clock, $\hbar \omega n$,
Eq. (\ref{as2}) provides a set of $\nu_n$, 
\begin{equation}
\label{mod}
\frac{\hbar \nu_n}{{\sf T}}
=\hbox{Mod}[\hbar \omega n, \frac{2 \pi \hbar}{{\sf T}}].
\end{equation}
Provided  
\begin{equation}
\label{omega2}
\hbar \omega j > \frac{2 \pi \hbar}{{\sf T}},
\end{equation}
or equivalently, $T>(2j+1)\tau/j$, 
and if the clock is to produce a small disturbance to the system,
the incident energy of the particle $E$ must satisfy
\begin{equation}\label{as1}E>> \lbrace
\hbox{Maximum value of  }
\frac{\hbar \nu_n}{\sf T}, \forall n  \rbrace=
\bigg( \frac{\hbar \nu_n}{{\sf T}}\bigg)_{max},
\end{equation}
but, by definition of $\nu_n$,
\begin{equation}\bigg( \frac{\hbar \nu_n}{{\sf T}}\bigg)_{max}
\le \frac{2 \pi \hbar}{{\sf T}},
\end{equation}
which together with Eq. (\ref{as1}) leads to
\begin{equation}\label{dsm}E>>\frac{2 \pi \hbar}{{\sf T}}.
\end{equation}
However, for sufficiently small ${\sf T}$ the kick clock behaviour 
resembles the one of the continuous-coupling clock and 
Eq. (\ref{peres}) holds instead of Eq. (\ref{dsm}).  
When
\begin{equation}\label{omega1}\hbar \omega j <
\frac{2 \pi \hbar}{{\sf T}},
\end{equation}%
then,
\begin{equation}
\hbox{Mod}[\hbar \omega j,\frac{2 \pi \hbar}{{\sf T}}]=
\hbar \omega n,
\end{equation}
and Eq. (\ref{peres}) is recovered.  
Since we must also have 
${t_f}>{\sf T}$,  
the proper working regime of the pulsed apparatus
is defined by the conditions
\begin{equation}
\label{interval}
{t_f}>{\sf T}>\frac{2j+1}{j}\tau.
\end{equation}
We may in addition set $2\pi/\omega$ as the maximum time to be
measured by the apparatus, to avoid the possibility 
of multiple times corresponding to a single $\theta$.
A measurement of $\theta$ at an asymptotically large
time well after
the particle-clock 
interaction will not tell us the number of $2\pi$ cycles
that have occurred,
so that the time read is only
known modulo $2\pi/\omega$.
This ambiguity may be avoided by substituting the
periodic hand motion by a linear one as in \cite{AOPRU98}. 
 
From our previous 
considerations it is clear that  the energy of the particle may violate  
the inequality in Eq. (\ref{peres}) and 
still lead to a succesfull time-of-flight
measurement for $T>(2j+1)\tau/j$, as we shall illustrate below with numerical 
examples in which   
a minimun uncertainty 
product Gaussian wave packet with negligible negative 
momenta is prepared at $t=0$ outside the 
region where the clock (continuous or kicked) is activated.
Of course, because of the momentum width
we should not expect a single time but a distribution.  
Well after the packet collision with the interaction region
the probability to find the value $\theta$ is calculated 
and the corresponding (operational) time of flight
distribution is obtained from the scaling $t=\theta/\omega$.  

Fig. 1A shows the ideal distribution of flight times
obtained for the system in isolation, ${\cal P}_d$, and the operational distributions
obtained from a kicked clock and from the continuous clock. 
(Incidentally, note that the continuous-coupling-clock distribution 
may also be obtained using D. Sokolovski's Feynman-path based theory \cite{Soko}).
The parameters are chosen so that the classical time of flight for
the average momentum of the wave packet 
is $10$ a.u., and in such a way that  
the inequality in Eq. (\ref{peres}) 
is not obeyed, i.e., the continuous-coupling clock does not 
work correctly:    
note the large early peak denoting an important reflection in its
distribution, and the displacement 
of the second peak with respect to ${\cal P}_d$
to shorter times because of the filtering 
effect of the more energetic barriers and wells
that hinder the passage of slower 
components and allows a dominant contribution of faster
components \cite{AOPRU98}.

The reference (ideal) curve ${\cal P}_d(t)$ is the distribution of the 
free-particle probability distribution
of dwell times.
For positive-momentum states 
the dwell time probability distribution is given by   
\beq 
{\cal{P}}_d(t)=\int_0^\infty dp\, \delta(t-md/p) P(p), 
\eeq
where $P(p)$ is the momentum distribution. ${\cal P}_d$ is both 
the dwell time distribution for 
an ensemble of classical particles with momentum distribution $P(p)$,
and the 
quantum dwell time distribution $\la \delta(t-\wh{\tau}_d)\ra$, 
where 
\beq
\wh{\tau}_d=\int_{-\infty}^{\infty} dt\,e^{i\wh{H}_st/\hbar} 
\left(\int_0^d |x\ra dx \la x|\right) e^{-i\wh{H}_st/\hbar}
\eeq
is the dwell-time operator \cite{Muga}.

Figs. 1B and 1C show the cumulative distributions for several values of   
${\sf T}$ and the cumulative distributions for 
${\cal P}_d$ and for the continuous-coupling clock.
As ${\sf T}$ is increased there is a passage from the  
continuous-like regime to the truly kicked regime,
where the cumulative distributions 
reproduce in a step-like fashion the behaviour of the 
reference ideal curve. The perturbation of the kicks may be seen in the 
broader wings, which grow with decreasing ${\sf T}$.

In summary, it is possible to extend the energy domain where 
a clock coupled to the particle's motion provides 
its (free motion) time-of-flight by using a pulsed 
particle-clock coupling rather than a continuous one.

\acknowledgments{The authors thank S. Brouard and
C.R. Leavens for many discussions. 
Support has been provided by Gobierno de Canarias (PI2000/111), 
Ministerio de Ciencia y Tecnolog\'\i a (BFM2001-3349 
y BFM2000-0816-C03-03), UPV-EHU (00039.310-13507/2001),
and the Basque Government (PI-1999-28).}


%

\begin{figure}
{\includegraphics[width=3.35in]{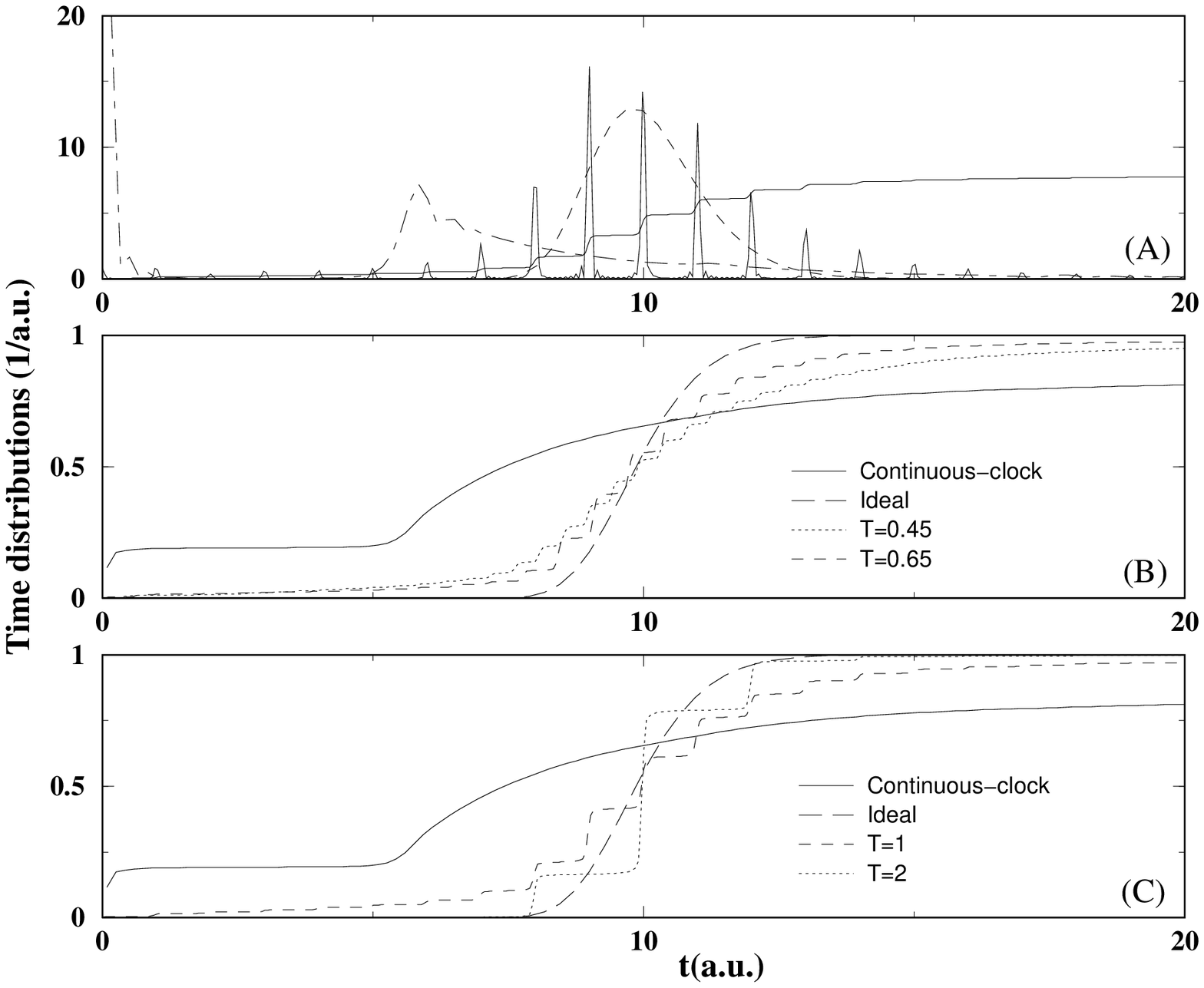}}
\caption[]{(A):  in solid line a typical ${\cal{P}}_d(t)$ and a cumulative
integration for $\sf{T}=1$. In dashed line the ideal time
distribution, Eq. (28), and in dotted-dashed line the time
distribution obtained from the continuous-coupling clock.

(B,C): Cumulative distributions for different values of $\sf{T}$
(dotted and dashed lines);   
cumulative
distribution for the continuous-coupling clock (solid
line) and for the ideal dwell time distribution (long-dashed line).

	The simulations were performed for a particle of mass $m=1$ a.u.
represented initially 
by a minimun-uncertainty-product wave packet with width $\sigma=1$ a.u.,
center at $x_0=-30$ a.u., and average momentum $p_0=5.0$ a.u..
The collision region
is the interval $x \in (-25,25)$ a.u.. The method used was a Split Operator
Method and we took $2^{13}$ plane waves for the spatial coordinate and
$2^{10}$ plane waves for the angular coordinate.}
\end{figure}

\end{document}